# Confronting Contemporary Seasonality Changes in East Asian Tropical Cyclone Landfalls with a Multi-Century Historical Baseline

Gan Zhang[1]*, Kuanhui Elaine Lin[2], Dan Fu[3], Tom Knutson[4], Jörg Franke[5], Wan-Ling Tseng[6]

[1] Department of Climate, Meteorology, and Atmospheric Sciences, University of Illinois at Urbana-Champaign

[2] Graduate Institute of Sustainability Management and Environmental Education, National Taiwan Normal University

[3] Department of Atmospheric Sciences, Texas A&M University

[4] Geophysical Fluid Dynamics Laboratory / National Oceanic and Atmospheric Administration

[5] Institute of Geography & Oeschger Centre for Climate Change Research, University of Bern

[6] Ocean Center, National Taiwan University

*Corresponding Author: Gan Zhang, gzhang13@illinois.edu

**Abstract**

**Paleoclimate records provide a critical long-term perspective on natural climate variability, essential for understanding contemporary climate change. However, existing paleoclimate proxies lack the spatial-temporal coverage for studying changes in high-impact weather extremes like tropical cyclones (TCs). Here we introduce a multi-source framework that confronts the contemporary changes in TC landfalls in East Asia with a multi-century baseline (1368–1911) reconstructed from historical documents. Leveraging pre-industrial and contemporary data, the analysis reveals that a relatively small shift toward earlier landfalls in the contemporary era (1946–2020). However, this shift falls well within the range of natural fluctuations documented historically (1651–1900). This low signal-to-noise ratio indicates the forced anthropogenic signal of TC landfall timing remains challenging to detect. Besides providing a template for assessing seasonality changes in extremes, our work shows consistent natural controls of TC timing in contemporary and pre-industrial eras, lending credibility to pre-industrial observational datasets and climate simulations.**

## Introduction

Tropical cyclones (TCs) are among the most destructive natural hazards, and understanding how their behavior may change in a warming world is a critical scientific and societal priority [1,2]. While much research has focused on changes in TC frequency and hazard intensity [3], shifts in their seasonality—the timing and duration of the TC season—can also have profound and distinct impacts [4,5]. An earlier or later TC season can alter regional rainfall patterns, compound monsoon-related flood risks, and challenge existing frameworks for disaster preparedness. Recent studies of the instrumental record have indeed suggested that the seasonality of TCs may be shifting [4,6,7]. However, these findings conflict with at least some climate model simulations [5,8], and the role of anthropogenic forcing remains uncertain.

A fundamental challenge in attributing observed changes in regional climate phenomena to anthropogenic forcing is disentangling the forced "signal" from the inherent "noise" of internal climate variability [9]. The modern instrumental record of tropical cyclones [10], spanning from the mid 1940s to the 2020s, provides only a single realization of decadal-to-centennial climate fluctuations. Furthermore, the cyclone counts are not reliable until satellite coverage improves in the 1970s; comprehensive estimates of cyclone hazard intensity were not available until later. Such short record lengths may be insufficient to determine whether an observed multi-decadal trend is a robust response to external forcing (e.g., human activities) or a transient feature of natural variability. A longer-term perspective is therefore essential to characterize the full range of natural variations and provide a baseline against which contemporary changes can be rigorously assessed.

Disentangling the drivers of TC activity changes also requires a reliable baseline without human influences. While climate models can simulate pre-industrial conditions to provide such a baseline, the accuracy of these simulations in representing pre-industrial TC activity remains a subject of active research [1]. A valuable data source for validating climate simulations is the TC reconstructions from coastal sediments that extend back centuries or millennia [11–14]. Recently, Yang et al. [15] compared the Atlantic TC frequency in the last millennium using sediment reconstructions and climate simulations. Besides showing the consistency of the decadal-to-centennial variations in sediment reconstructions and climate simulations, the study also suggested that large last-millennium variations in the Atlantic TC frequency were driven by the natural variability of the climate system. Despite the encouraging results, sediment-based TC reconstructions lack the temporal resolution needed to serve as a baseline for intra-decadal variations in TC activity.

To construct this long-term perspective, we turn to the Reconstructed East Asian Climate Historical Encoded Series (REACHES), a unique dataset compiled from Chinese documentary records spanning from 1368 to 1911 [16–19]. While this archive has inherent limitations, including potential biases from societal exposure, changing documentation practices, and inconsistent event classification (see Supplementary Materials), it is invaluable for its multi-century depth and potential of independently validating other observational datasets and climate simulations. Since the REACHES documentation of cyclone intensity and regional frequency is only qualitatively informative [20], this study focuses on analyzing the timing of TC landfalls. This metric, especially when aggregated over multiple decades, is likely quantitively reliable thanks to the calendar system updates. To formally confront contemporary trends with this historical baseline, we integrate the REACHES data with modern instrumental records (IBTrACS, 1946–2020) [10] and high-resolution climate model simulations (CESM-HR) [21].

Numerous studies have used contemporary observations and climate simulations to link spatial-temporal variations of TC activity to large-scale climate variability[5,22–26]. However, paleoclimate studies often had to interpret changes in TC activity with site-based proxy climate data [27], an approach that can be problematic because a limited number of sites may not accurately represent broader regional or global climate patterns. To overcome the limitations of relying solely on site-based proxies, we leverage the recently developed ModE-RA dataset, a spatially comprehensive global monthly paleo-reanalysis [28]. ModE-RA constrains an ensemble of climate simulations by assimilating a wide range of paleoclimate data. Because the assimilated data does not include the REACHES TC records, the two datasets remain independent of each other. By linking variations in these two datasets and comparing the relationship with its counterparts in other climate periods, we can gain valuable insights into the potential climate drivers of long-term TC variability.

With this multi-century historical context, we confront contemporary changes in the seasonality of East Asian TC landfalls. Our analysis suggests these recently observed shifts are not statistically distinguishable from the range of natural variability documented in the pre-industrial era. The work also develops and demonstrates a comprehensive framework that validates extreme events in paleoclimate records and climate simulations. The paper is structured as follows: First, we characterize the climatology and multi-century variability of TC seasonality and landfall latitude within the REACHES record to establish a robust baseline of pre-industrial behavior. Second, we compare trends from the contemporary observational and modeling era against this baseline to evaluate their significance and novelty. Finally, we explore the large-scale climate patterns associated with historical seasonality shifts to diagnose their potential physical drivers. This multi-source approach provides a crucial long-term context for interpreting contemporary TC trends and the formidable challenges of their attribution.

## Results

### Climatology and Long-Term Variability in Historical Records

We begin by characterizing the pre-industrial baseline of TC activity using the Reconstructed East Asian Climate Historical Encoded Series (REACHES) [19,29], a dataset compiled from Chinese documentary records from 1368–1911 (Methods). The number of TC records increases substantially over this period, particularly after the transition from Ming (1368–1644) to Qing (1644–1911) dynasty (Figure 1a). The year 1644 witnessed the collapse of the Ming dynasty, which was weakened by chronic fiscal stress and widespread peasant revolts associated with severe natural disasters, including droughts, floods, and famines [30,31]. The low number of TC records before the mid 1600s may result from the destruction by peasant wars and Qing dynasty. Meanwhile, the substantial population growth in China from ~65 million to ~450 million likely increased societal exposure to TC hazards, potentially leading to more frequent and systematic documentation of these events. The large increase of TC records between 1650 and 1700 may also indicate physical changes partly driven by elevated TC activity linked to the solar activity [19]. Amidst these profound societal and natural shifts, Chinese intellectuals undertook calendar reforms around 1650s by incorporating the astronomy knowledge from Western missionaries. These updates made the Chinese and Western calendars comparably reliable, lending confidence to the date record of TC landfall after 1650.

To assess the quality of these historical TC records, we first perform a consistency check of their broad climatological features against the contemporary instrumental era (1946–2020; Best

Track [10]). Figure 1b shows the monthly distributions of TC records in the REACHES and the contemporary observational data. Both datasets show a consistent concentration of landfall TCs between June and November, peaking around August–September. Meanwhile, the latitudinal range of REACHES TC records is also consistent with the contemporary-era data (Figure 1c), though the REACHES records show a greater concentration near 23ºN and 31ºN. These latitudes correspond to the vicinities of major seaports in the Pearl River Delta (e.g., Guangzhou in South China) and the Yangtze River Delta (e.g., Shanghai in East China), respectively. While potentially influenced by observational biases, the broad spatial-temporal consistency with contemporary data suggests the historical records plausibly capture the fundamental seasonality and spatial patterns of TC landfalls in East Asia.

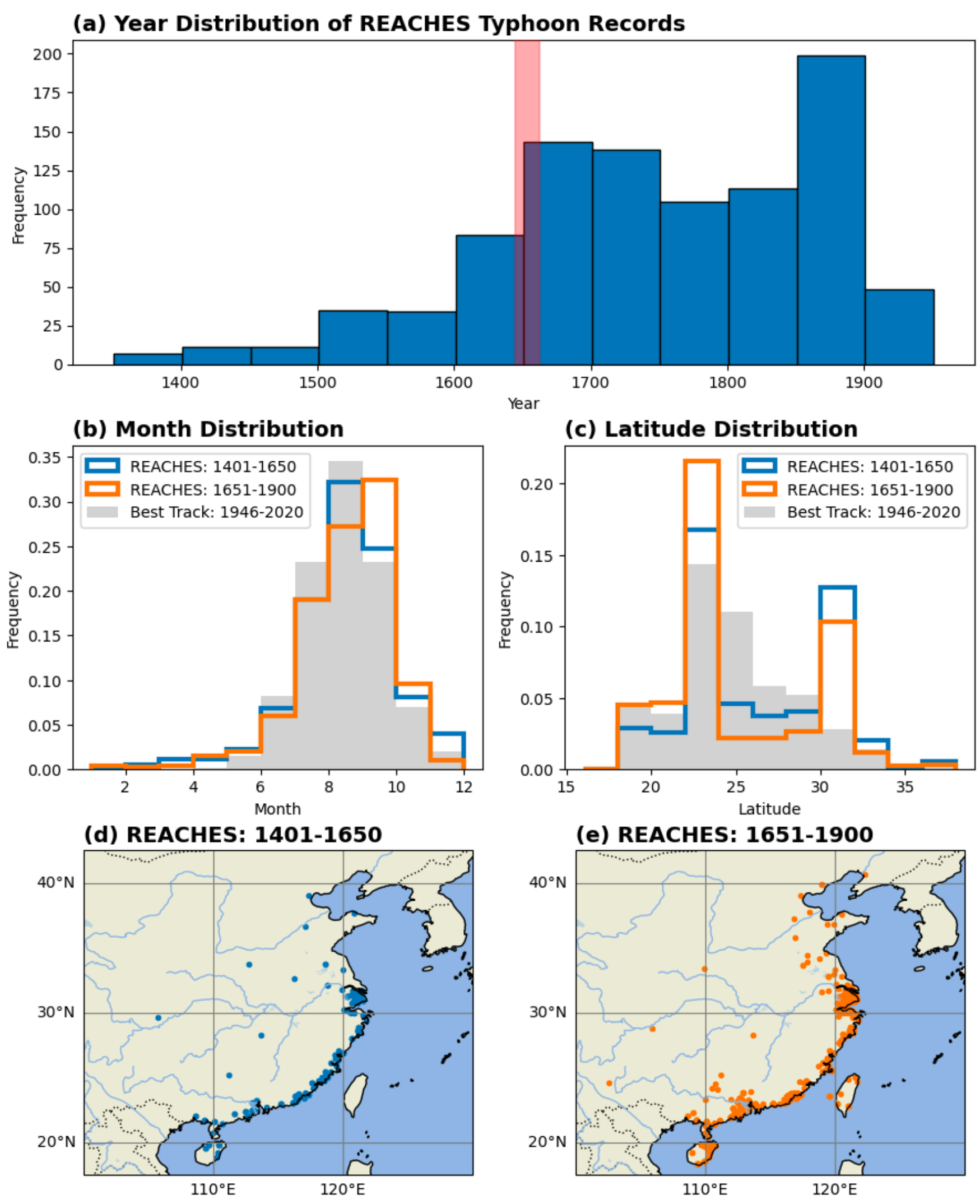

*Figure 1 Overview of the REACHES TC records. (a) REACHES record counts in 50-year aggregations during 1368–1911. The red shading in a) highlights the transition period between the Ming and Qing dynasties. (b) Monthly distributions of the REACHES during 1401–1650 (blue) and 1651–1900 (Orange), as well as the contemporary-era best track data (1946–2020). The distributions are scaled with the total storm counts in the corresponding datasets. (c) Same as (b), but for the latitudinal distributions. (d) The spatial distribution of the REACHES TC records during 1401–1650. (e) Same as (d), but for 1651–1900.*

Beyond this climatology, the REACHES record reveals pronounced multi-century variations in TC landfall timing (Fig. 2a) and location (Fig. 2b). To quantify long-term variations, we analyze the REACHES TC records using 50-year moving windows (Methods). The day of year (DoY) of TC landfalls reveals notable variations over the period of 1401–1900, including variations as large as 20 days in the 50-year mean DoY. Notably, the season onset ($5^{th}$ percentile DoY) varies by nearly 100 days over the same period. Meanwhile, Figure 2b reveals a southward shift in the mean latitude of TC records after the late 1500s, from approximately 28°N to 23°N. While this southward shift could be physical [18,32–34] and is consistent with drought conditions in North and Central China [30], data limitations (e.g., heterogeneous record-keeping) and other issues complicate the interpretation of this latitudinal shift (see Supplementary Materials). Nonetheless, the substantial seasonality fluctuations, occurring in the absence of significant anthropogenic forcing, provide a crucial, upper-bound estimate of the magnitude of natural variability against which contemporary changes must be evaluated.

## Confronting Contemporary Trends with Historical Variability

We next formally confront this historical baseline with the contemporary changes, as well as high-resolution climate simulations (Figs. 2c and 2d). For contemporary-era observations, we analyze landfalling TCs in 1946–2020 and evaluate the 75-year statistics. This long period is selected to maximize the use of modern observations after World War II. We also examine the high-resolution Community Earth System Model simulations (CESM-HR; Methods) [21], which provide a physically-based representation of TC activity under different climate conditions. We assume that I) the climate states in the REACHES and the CESM-HR pre-industrial control experiment are broadly similar and can represent natural variability in the absence of significant anthropogenic forcings; II) the climate states captured by the contemporary observational data and the CESM-HR historical experiment are comparable and can realistically represent the natural variability and growing anthropogenic forcings; and III) comparing the observed and simulated changes helps evaluate the robustness of potential changes associated with anthropogenic warming.

We first note that the DoY and latitude metrics from the REACHES records and recent observations are generally consistent with the overland data in the CESM-HR simulations. The most notable intra-group discrepancy in Figures 2c–d is associated with the 75$^{th}$ and the 95$^{th}$ percentile values of TC latitude in the REACHES records. This discrepancy may in part stem from the high concentration of TC records near Shanghai relative to adjacent regions (Figure 1), suggesting the statistics of latitude shift in REACHES can be sensitive to observational limitations. In contrast, the DoY differences among the two observational datasets and the CESM simulations are much smaller, suggesting that landfall timing is more reliably recorded than the landfall location. These results suggest that comparing observed and simulated changes in TC landfall timing is more feasible.

A direct comparison of TC landfall timing between the late pre-industrial (1776–1850; REACHES) and contemporary (1946–2020; Best Track) periods reveals a mean shift of approximately one day toward earlier landfalls in the observational datasets (Fig. 2c). While this shift is not statistically significant ($p > 0.1$; Kolmogorov-Smirnov test), the result is highly sensitive to the chosen analysis window. For example, comparing 1801–1850 of REACHES records with 1971–2020 of Best Track recrods yields a small yet significant seasonal advance ($p < 0.01$; not shown). In contrast to the observed shift, the CESM-HR model simulates a relatively large delay (~5 days; CESM-hist minus CESM-ctl) in TC seasonality over the same period. The delay is qualitatively consistent with from other climate model simulations [5,8], highlighting a model-observation discrepancy. Interestingly, the CESM-HR simulations also show that the statistical significance of the simulated DoY changes depends on the analysis period, with weaker statistical significance for DoY differences when comparing longer 75-year periods (p-value<0.1) than shorter 50-year periods (p-value<0.05; not shown).

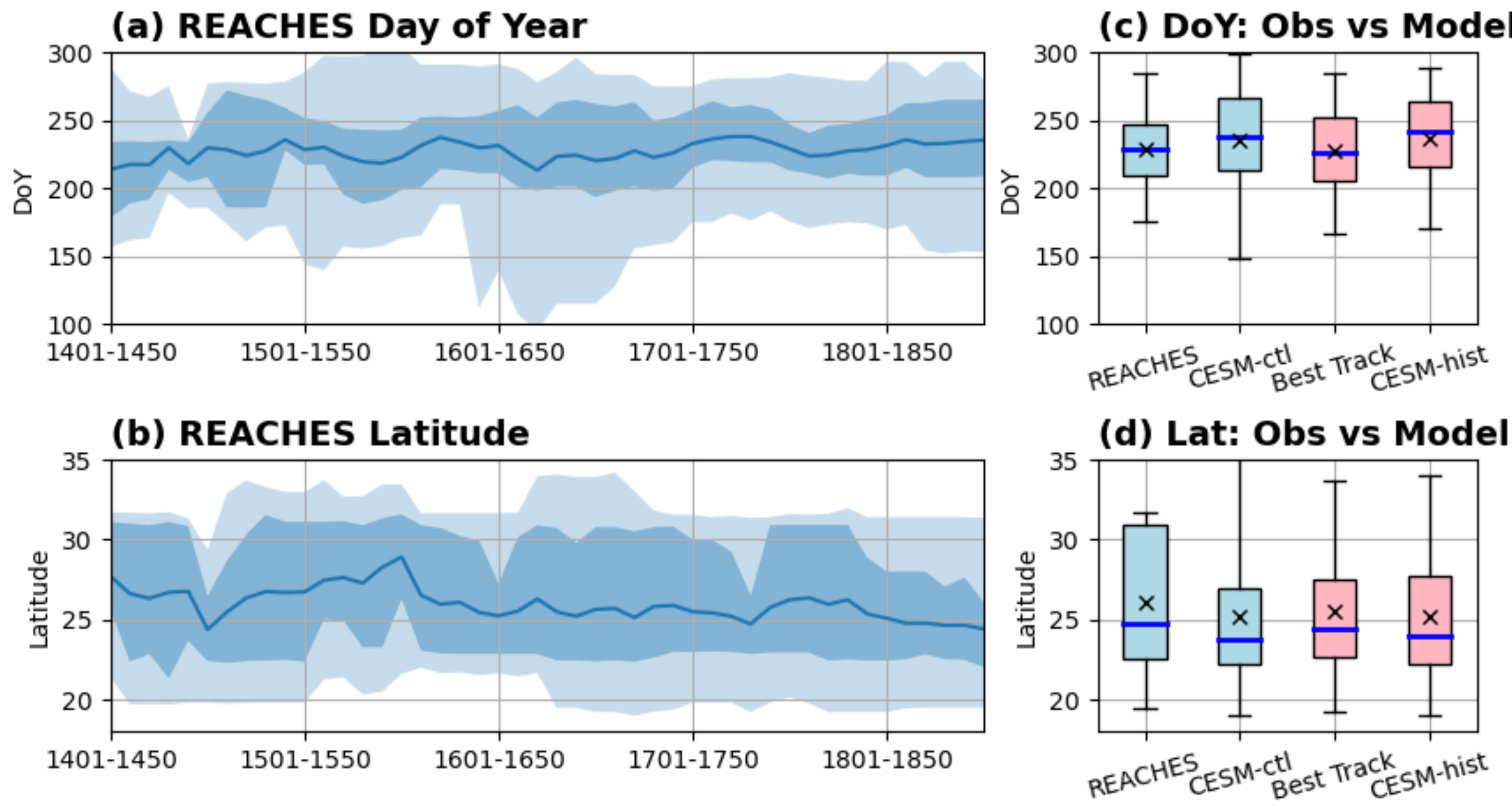

*Figure 2 Long-term changes in the spatial-temporal distributions of REACHES TC records and the snapshot comparison with other datasets. (a) The evolution of 50-year aggregation statistics of the day of the year (DoY) of the REACHES TC records during 1401–1900. The dark blue line shows the mean DoY. The deep and light shading denote the 25$^{th}$-75$^{th}$ percentile range and the 5$^{th}$-95$^{th}$ percentile range, respectively. (b) Same as (a), but for the latitudes of the REACHES TC records. (c) The statistics of the DoY of TC data in the REACHES (1776–1850, light blue), the CESM-HR control experiment (1850-scenario, light blue; a randomly selected 75-year segment), the recent observational data (1946–2020, light red), and the CESM-HR historical experiment (1946–2020, light red). The boxplots in denote the 5$^{th}$, 25$^{th}$, 50$^{th}$ (blue line), 75$^{th}$, and 95$^{th}$ percentile values of the sample populations. The crosses show the means of each sample. (d) Same as (c), but for the latitude of TC records.*

The sensitivity to the analysis period underscores the challenge of isolating a robust, forced trend from a record characterized by large internal variability. The higher statistical significance associated with comparing 1801–1850 and 1971–2020 has at least two interpretations. One possible interpretation is that the human-forced changes are stronger and more detectable in the late twentieth century. However, this interpretation cannot explain why the DoY shifts in the observations (Best Track minus REACHES) and simulations (CESM-hist minus CESM-ctl) have opposite signs (Figure 2c). Attributing these shifts to anthropogenic forcings would need strong scientific evidence that can invalidate the observational datasets or the CESM-HR and other simulations [5,8]. An alternative interpretation is that natural variability drives landfall timing variations larger than the human-driven signals, making it challenging to reliably identify human-driven signals using a single, relatively short climate realization.

A useful assessment of the second interpretation is to compare the magnitude of the 75-year contemporary trend to the full distribution of trends that occurred naturally in the past (Figure 3). Figures 3a and 3b show trends calculated over a range of multi-decadal periods, illustrating how the detection of statistically significant trends can be influenced by the choice of analysis time window. The statistically significant trends in the pre-industrial periods may arise from unusual natural climate forcing (e.g., solar radiation and volcanic activities), unknown record-keeping anomalies, or other method uncertainties. With this context acknowledged, we directly confront the trend from the IBTrACS record with the histogram of all possible 75-year trends derived from the REACHES dataset (Figure 3c). The magnitudes of the 75-year trends in DoY and latitude in the IBTrACS and CESM historical simulations are close to zero. Furthermore, the Hamed-Rao modified Mann-Kendal test (Methods) suggest these trends are statistically insignificant. Lastly, the observed contemporary trend falls squarely within the 95% confidence interval of the historical distribution, indicating that this trend is not unprecedented.

Based on this confrontation of observed and simulated changes, we adopt the framework of Ref [9] and classify the situation as Uncertain. This designation is warranted when a definitive conclusion about model performance is prevented by fundamental challenges because: the observed trend is not strong enough to be distinguished from the "noise" of natural climate variability, or because significant flaws or biases in the observational record make it an unreliable benchmark for model evaluation. Here, both sources of uncertainty are relevant. Therefore, a definitive conclusion on whether the model is "right" or "wrong" cannot yet be drawn for this specific regional metric, and attributing the observed trends in landfall timing solely to human influences is premature.

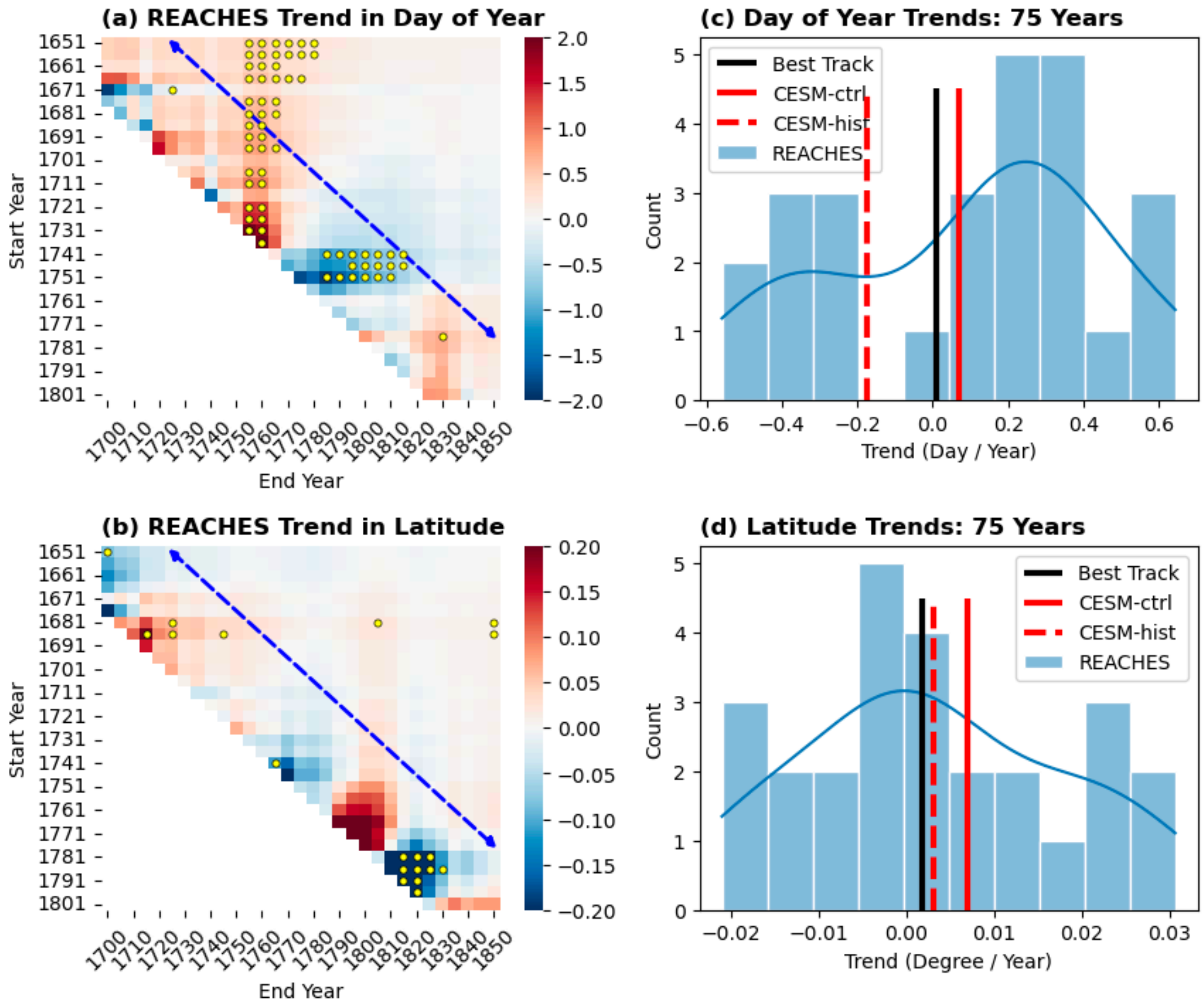

Figure 3 *Comparison of trends in the REACHES data and other datasets. (a) Heatmap of the trends in the annual means of the day of the year of the REACHES TC records. The vertical axis indicates the starting year of analysis segments, and the horizontal axis indicates the end year. The trends above the 95%-confidence level based on the modified Mann-Kendal tests (Methods) are indicated by yellow dots. The blue dashed line highlights the trend values of the 75-year REACHES segments displayed in the right subplot (1651–1725, 1656–1730, and so on). (b) Same as (a), but for the annual mean latitudes of the REACHES TC records. (c) The histogram of the 75-year trend values of the annual mean day of the year, with the estimated kernel density denoted by the blue line. The analysis focuses on the 1651–1900 period, which has relatively abundant REACHES TC records. The black line indicates the trend value in the recent observational data ("Best Track"). The red solid and dashed lines indicate the trend value of the CESM-HR control experiment (1850-scenario, randomly selected 75-year segment) and the CESM-HR historical experiment (1946–2020), respectively.*

## Potential Climatic Drivers of Seasonality Shifts

Understanding the physical drivers of the multi-century seasonality variations in the REACHES record requires linking them to the evolution of the large-scale climate. While studies of recent and future TC activity often leverage atmospheric reanalysis data to investigate climate controls, such data is less reliable and offers few variables for the pre-industrial era. Therefore, the analysis here focuses on the sea surface temperature (SST), which interact with the atmosphere and has been linked to several mechanisms that influence TC timing.

Recent studies identified distinct effects of local, remote, and global SST changes on TC seasons. Local SST warming has been argued to advance TC-favoring thermodynamical environment and drive an earlier onset of the TC season in the Atlantic [6,7] and for intense TCs in the Northwest Pacific [4]. Nonetheless, the precise pattern of local SST changes in the Northwest Pacific can complicate TC responses [8]. Besides the local SST, the importance of remote SST was highlighted by observational research. Specifically, strong La Niña condition suppresses early-season TC activity in the Northwest Pacific, and vice versa [23]. Coincidently, the observed tropical Pacific trend over recent four decades has been La Niña-like [2,35]. If the observational relationship in the contemporary era holds [23], this trend might have delayed TC timing and counteracted the effect of the regional warming of Northwest Pacific. Lastly, climate models forced with $CO_2$ and other anthropogenic forcings robustly project *global* SST warming. The warming leads to a delayed seasonal migration of tropical precipitation [36,37], which has been linked to a delay of TC season [5,8]. Regardless of whether the effects of regional, remote, or global SST changes dominate the landfall timing in East Asia, it is valuable to assess and understand these SST-TC relationship in a longer climate context.

Accordingly, we correlate the historical TC records with the surface temperature from the ModE-RA paleo-reanalysis (Methods) (Figure 4). The analysis reveals a robust link between delayed TC landfalls (higher DoY) and a distinct climate pattern. Spatially, this pattern features La Niña-like temperature in the tropical Pacific, as well as warmer-than-average conditions over mid-latitude East Asia and Northwest Pacific (Fig. 4a, 4c). The broad pattern of La Niña-like signals and mid-latitude warming and their impacts on the seasonality of TC activity is qualitatively consistent with the findings of observational [23] and modeling [8] studies of the recent and future climates. Therefore, this finding strengthens the credibility of both the REACHES TC records and the ModE-RA reanalysis, as their independent variations are linked by physically plausible mechanisms.

Robust precipitation-TC relationship (Figure 4b, 4d) is also present and consistent with variations in surface temperature and monsoon. Specifically, the wetting near Indonesia and the drying in the Central Pacific are robustly linked to late TC timing (Figure 4d). The precipitation pattern is consistent with the tropical Pacific SST forcing related to La Niña-like signals. A delayed DoY (i.e., later TC landfalls) is significantly correlated with negative precipitation anomalies (i.e., drier conditions) around July, primarily around 10ºN and 35ºN (Figure 4b). Dry signals also appear in the Bay of Bengal and the South China Sea (Figure 4d), suggesting an association between the Asian monsoon and the timing of TC landfalls.

Most of the climate-TC relationships identified using REACHES and ModE-RA are corroborated by an independent reanalysis dataset [38] (20CRv3; Methods) during 1811–1900 (Supplementary Figure 3) and contemporary era (Supplementary Figure 4). For instance, a delayed DoY in other climate periods and datasets is also linked to La Niña-like surface temperature anomalies (Supplementary Figures 3 and 4). While disagreements in pattern details exist, the overall agreement on the link between the El Niño-Southern Oscillation (ENSO), DoY, and precipitation anomalies strengthens the credibility of the pre-industrial findings.

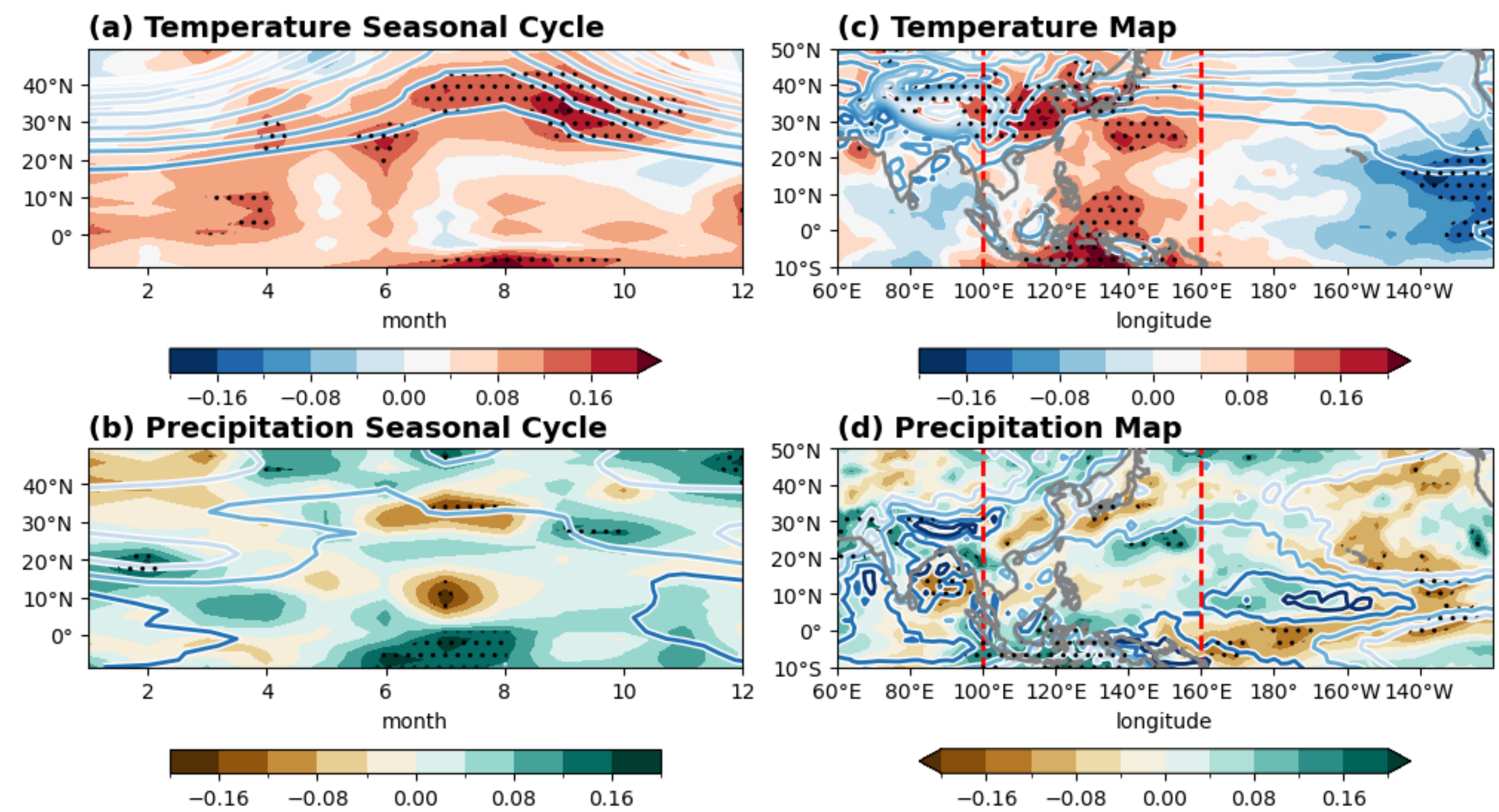

*Figure 4 Correlations between the annual mean DoY of the REACHES TC records and the ModE-RA variables. (a) Correlation with the monthly mean, zonal mean surface air temperature (shading) during 1651–1900. The temperature climatology (1901–2000) is denoted by contour lines. The range of zonal averaging is 100ºE–160ºE. (b) Same as (a), but for the precipitation. (c) Correlation with the June–October means of surface air temperature (shading). The annual mean DoY is raw yearly data without any smoothing. The climatology (1901–2000) is denoted contour lines. The red dashed lines highlight the zonal averaging range in (a). (d) Same as (c), but for the precipitation. The stippling highlights the signals above the 90% confidence level.*

## Discussion

Our analysis leverages multi-century historical documents and newly available paleoclimate reanalysis to provide a crucial long-term context for interpreting contemporary trends in the seasonality of East Asian TC landfall. The primary finding is that while a subtle shift toward earlier landfalls may be occurring in the contemporary era, the anthropogenic signal in this regional metric has not yet robustly emerged from the noise of natural climate variability. The magnitude of the contemporary 75-year trend is not statistically unusual when compared to the distribution of trends that occurred naturally during the pre-industrial period (1651–1900). This highlights the formidable challenge of detecting and attributing changes in regional TC characteristics against a backdrop of high intrinsic variability.

The simulated changes in the Northwestern Pacific TC landfalls are classified as Uncertain, which stems from a combination of factors outlined by the framework for model-observation confrontation [9]. First and foremost is the low signal-to-noise ratio. The large, multi-decadal fluctuations in TC timing within the REACHES record (Fig. 2a) establish a high threshold of natural variability that any forced signal must exceed to be detectable. Second, we identify a model-observation discrepancy. The CESM-HR model simulates a seasonal delay relative to the pre-industrial era, which is opposite to the earlier onset indicated by observational data. This model behavior may be linked to well-documented large-scale biases, such as the precipitation biases in the Asian monsoon region and the tropical Pacific [21].

An encouraging finding is the consistency between the spatial-temporal variations of TC activity derived from the REACHES dataset and those in contemporary instrumental records. Significant correlations were identified between the long-term variations in the REACHES TC records and large-scale climate patterns in the ModE-RA paleoclimate dataset. The relationship is corroborated by additional climate datasets and broadly consistent with recent studies of different periods. These findings bolster confidence in the reliability of TC records in the REACHES and the ModE-RA dataset. Together they suggest that the strong seasonal timing variations in the pre-industrial TC records are shaped by natural climate variability, rather than being solely attributable to societal changes or data artifacts in the pre-industrial era.

We acknowledge several inherent limitations of our study. The REACHES TC records, while providing unparalleled temporal resolution and length, lack intensity information and are subject to documentation biases that cannot be fully resolved. Furthermore, our analysis of climate drivers is limited by variable availability and relies on statistical correlations, and a deeper mechanistic understanding is warranted. Future research should prioritize a direct analysis of TC activity in high-resolution climate models to determine whether the observed seasonality shifts are driven by changes in genesis location or track shape. Such analyses, combined with single-forcing experiments to isolate the effects of greenhouse gases [39], would be crucial for moving from detection to a confident attribution of changes in seasonality and other aspects of TC activity.

Confronting observed and simulated changes in TC activity is valuable yet challenging [9], especially in contexts involving the pre-industrial era that lacks the reliable, comprehensive instrument observations. The agreement between REACHES and independent paleoclimate reanalysis—corroborated by modern climate data—highlights the potential of integrating diverse data sources to unravel the complex interplay of natural variability and anthropogenic forcing on TC activity and other extremes.

## Methods

### TC Records in Historical Documents

This study analyzes the TC records from the Reconstructed East Asian Climate Historical Encoded Series (REACHES) [19,29]. The REACHES dataset digitizes textual descriptions, dates, and locations of significant weather and climate events documented by the "Compendium of Meteorological Records of China in the Last 3000 Years" [17]. This comprehensive compendium compiles weather and climate records from a vast collection of chronicles and other historical documents. We extract the TC records in the REACHES by searching for Chinese words corresponding to TCs, which represent the historical, crowd-sourced understanding of TCs by intellectuals in pre-industrial China. To minimize potential record uncertainties, we focus on land-based records and exclude the data records without a clear date (e.g., year information only). We refer readers interested in the documentation and validation of the REACHES TC records to Refs *22*, *24*. The spatial-temporal variations of TC activity in the pre-1850 records can be interpreted as responses to societal exposure changes (e.g., population migration), natural climate variability, and limited early anthropogenic influences (e.g., land use change). A key limitation of the TC records in the REACHES dataset is the lack of TC intensity information and clear associations with non-wind hazards. Details about these limitations of the dataset are available in Supplementary Materials.

### Contemporary-era Observations and Simulations

To evaluate and contextualize the TC activity changes revealed by the REACHES dataset, we also analyze the TC tracks from the International Best Track Archive for Climate Stewardship (IBTrACS) [10] and the high-resolution Community Earth System Model (CESM) simulations [21] (CESM-HR). The IBTrACS is a state-of-the-art observational dataset of contemporary-era TC activity. This dataset consists of information from diverse observation platforms (e.g., satellites and airplanes) and is quality-controlled by experts. For consistency with the REACHES data, this study focuses on the overland storm data. We use the data of 1946–2020 from the U.S. Joint Typhoon Warning Center. During this period, changes in TC activity can be attributed to observation capability changes, natural climate variability, and anthropogenic influences. The analyzed CESM-HR simulations include the 1850-control experiments and the 1920–2100 historical experiment. The CESM-HR simulations employed a nominal horizontal resolution of 0.25° for the atmosphere and land components, and 0.1° for the ocean and sea-ice components. CESM-HR has produced a 500-year preindustrial control (PI-CTRL) climate simulation and a historical-and-future transient (HF-TNST) climate simulation from 1850 to 2100, branched from PI-CTRL at year 250, following the protocol for the Coupled Model Intercomparison Project phase 5 (CMIP5) experiments. PI-CTRL was forced by a perpetual climate forcing that corresponds to the 1850 conditions, while HF-TNST was forced by the observed climate forcing until 2005 after which the climate forcing follows RCP8.5 emission scenario. While changes in the 1850-control experiment result from internal natural variability, changes in the historical experiment result from internal natural variability and anthropogenic forcings. More details of these simulations are available in Chang et al. (2020). The TCs in the CESM-HR simulations are tracked using the TempestExtremes algorithm [40], so their differences from the IBTrACS TCs may arise from model or tracking biases.

### Reanalysis Datasets

We complement the analysis of TC activity in the REACHES by examining the historical climate represented by the Modern Era Reanalysis (ModE-RA) [28]. The ModE-RA is a monthly paleo-reanalysis spanning 1421 to 2008. Starting from an ensemble of simulations constrained by climate forcings, this dataset combines modelled climate fields with observations using an offline data assimilation approach. Assimilated observations include natural proxies and documentary data before the 17th century, and afterwards additionally instrumental measurements. Sparsely populated regions, including the tropical Pacific, are not strongly constrained by observations before the 19$^{th}$ century. Although the ModE-RA does not resolve TCs and has limited variables, it provides greater spatial and temporal coverage of the historical climate system than the site-based proxy data employed by previous paleoclimate studies of TC activity. With the caveats that ModE-RA has its own limitations and is a reconstruction, not a direct observation, of past climate states, we use the ModE-RA to explore potential connections between landfalling TC activity and variations in the large-scale climate (e.g., tropical precipitation) suggested by recent studies [4,5].

After identifying the climate-TC relationship using the ModE-RA and REACHES, we examine the sensitivity of the relationship to analysis periods and datasets (Supplementary Figures 4 and 5). The sensitivity tests leverage the IBTrACS and two additional reanalysis datasets, the NOAA Twentieth Century Reanalysis system version three (20CRv3) [38] and the fifth generation of the ECMWF Reanalysis (ERA5). The 20CR assimilates observations of surface pressure and sea surface temperatures and covers the period of 1806–2015. This extended temporal coverage helps evaluate the climate-TC relationship in different periods (e.g., 1811–1900 and 1946–2010), enabling comparisons with findings from other datasets. The ERA5 is state-of-the-art reanalysis that assimilates extensive observations and covers the period from 1940 to near present.

### Data Analyses

The investigation starts with a descriptive analysis of the REACHES TC records and then quantifies the spatial-temporal changes. The REACHES TC records span several centuries with varying levels of completeness (Figure 1 and Supplementary Figure 1). The descriptive analysis ("TC Climatology in Historical Documents") provides an overview of these TC records and their historical context, laying the foundation for the subsequent analyses. We then sample 50-year windows of the REACHES data for several reasons. First, aggregating data over 50 years helps extract robust TC statistics (e.g., seasonal cycle) during the periods with relatively scarce data. Second, the 50-year segment length enables a relatively large sample size that helps explore the sensitivity of identified climate trends to the specific analysis period. The analysis made an exception in the comparisons involving with the contemporary-era observational data, where we use 75-year segments to include more landfalling TCs available in the contemporary-era datasets. Within each multi-decade segment, we calculate key statistics (e.g., means and percentiles) and trends in the annual mean values of the day of the year (DoY) and the latitudes of TC records. After quantifying potential shifts in TC timing and location, we estimate the statistical significance of identified trends using the Hamed-Rao modified Mann-Kendal test [41], which is non-parametric method that detects monotonic trends while accounting for data autocorrelation. Lastly, we perform a correlation analysis to investigate the relationship between the variations of these annual mean TC metrics and the large-scale climate patterns represented by contemporary-era [42] and paleoclimate [28] reanalysis datasets. Additional details such as the analysis periods are available in the figure captions.

## Acknowledgments

G.Z. thanks Dr. Gregory Hakim for suggestions on the paleoclimate reanalysis and Drs. Renzhi Jing and Jie Chen for stimulating suggestions. The research is supported by the U.S. National Science Foundation award AGS-2327959 and the faculty development fund of the University of Illinois at Urbana-Champaign. D.F. acknowledges the funding supports from the U.S. National Science Foundation award AGS-2231237 and AGS-2332469. K.L. acknowledges the funding support from the Science and Technology Council of Taiwan NSTC 112-2122-M-001-001. J.F. is funded by the Swiss National Science Foundation grant number 219746.

## Author Contributions

G.Z. conceived the study and conducted the analyses. G.Z. drafted the manuscript with the input of all the authors. D.F., J.F., and K.L. contributed to the analyses and interpretation of CESM simulations, ModE-RA data, and REACHES records, respectively. T.K. and G.Z. developed the statistical analyses.

## Competing Interest Declaration

The authors declare no competing interests.

## Data Availability

The REACHES, IBTrACS, and ModE-RA datasets are archived by NCEI/NOAA. The archived data links are as follows: the REACHES (https://www.ncei.noaa.gov/access/metadata/landing-page/bin/iso?id=noaa-historical-23410), the IBTrACS (https://www.ncei.noaa.gov/products/international-best-track-archive), and the ModE-RA (https://www.ncei.noaa.gov/access/paleo-search/study/38239). NOAA/CIRES/DOE 20th Century Reanalysis (V3) data are provided by the NOAA PSL, Boulder, Colorado, USA, from their website at https://psl.noaa.gov. The CESM-HR data used in this work are available from https://ihesp.github.io/archive/products/ihesp-products/data-release/DataRelease_Phase2.html. The code used to generate the plots will be accessible via a Zenodo repository before publication.

Supplementary Materials for

# Confronting Contemporary Seasonality Changes in East Asian Tropical Cyclone Landfalls with a Multi-Century Historical Baseline

Gan Zhang[1]*, Kuanhui Lin[2], Dan Fu[3], Tom Knutson[4], Jörg Franke[5], Wan-Ling Tseng[6]

[1] Department of Climate, Meteorology, and Atmospheric Sciences, University of Illinois at Urbana-Champaign

[2] Graduate Institute of Sustainability Management and Environmental Education, National Taiwan Normal University

[3] Department of Atmospheric Sciences, Texas A&M University

[4] Geophysical Fluid Dynamics Laboratory / National Oceanic and Atmospheric Administration

[5] Institute of Geography & Oeschger Centre for Climate Change Research, University of Bern

[6] Ocean Center, National Taiwan University

## 1 Limitations of TC Records in REACHES

Using historical records like REACHES to reconstruct past TC activity comes with several inherent limitations. The capabilities and practices of human societies in observing and recording TCs differ substantially between the pre-industrial era and the contemporary era. The pre-industrial records have notable gaps in spatial (Figure 1) and temporal (Supplementary Figure 1) sampling. Over a short period of time (e.g., 10 years), the smaller number of samples may make the statistics noisier. Furthermore, the criteria used to identify and categorize events in REACHES do not perfectly align with current meteorological definitions. For example, the TC identification by pre-industrial intellectuals did not involve key quantities (e.g., maximum near-surface wind speed) or understanding (e.g., upper-tropospheric warm-core structure) of the contemporary TC definition. This also results in the lack of TC intensity information as defined by contemporary standards. Consequently, our analysis cannot distinguish between the spatial-temporal variations of TCs of different intensity. The lack of intensity information also limits the applications of REACHES for other topics of climate change, such as detecting the warming-induced changes in TC intensity. Lastly, the event classification may vary among record contributors across different times and regions. Some events documented in REACHES, such as instances of heavy rainfall and flooding, could have been associated with TCs or their broader synoptic influences, even if not explicitly categorized as TC-related (see Figure 8 in Ref [29]).

While these limitations preclude a direct one-to-one comparison with contemporary-era TC data, we emphasize that REACHES nonetheless provides valuable, albeit imperfect, insights into long-term variability of TC landfalls. Future research could explore methods to address these limitations, such as developing proxy indicators for TC intensity from associated event descriptions or pursue synergies with other datasets (e.g., sediment proxy data and storm downscaling from reanalysis data). Despite these challenges, the REACHES dataset remains a unique and valuable resource for extending our understanding of TCs and other extremes beyond the instrumental record.

## 2 Latitudinal Shifts in REACHES TC Records

The latitudinal shifts in the contemporary-era and pre-industrial periods are relatively hard to interpret. In the pre-industrial period, a southward shift of the mean latitude of REACHES TC records around 1550 (Figure 2b) is consistent with the previous analyses of TC report counts that used the same dataset [18]. Analyses of the recent sediment-based TC reconstructions also suggest a relative increase of intense TCs in South China during the same period [32–34]. However, the sparse data records before 1600 (<1 per year) and record concentration near Guangzhou after 1651 raise concerns about the nature of this southward shift. Other evidence also limited our confidence in interpreting the latitudinal shift. Specifically, the latitudinal shift during 1651–1900 mostly failed to pass the statistical significance test (Supplementary Figure 1b). While the latitudinal samples from the recent observational data appear significantly farther southward compared to the REACHES samples (Figure 2d), this metric in the CESM-HR control experiment and historical experiment are statistically indistinguishable (Kolmogorov-Smirnov test, $p\text{-value}>0.1$). The comparison of the pre-industrial and contemporary-era data (Figure 2) suggests inconsistencies among observational records and climate simulations, suggesting large errors in record-keeping, storm tracking, or the climate simulations. Compared with the DoY shift, the lack of multi-line evidence support weakens our confidence in the latitudinal shifts.

To further explore drivers of long-term changes in TC activity, we conduct correlation analyses similar to Figure 4 and examine the relationship between the annual mean latitude of TC records in REACHES and the ModE-RA climate fields (Supplementary Figure 2). The results reveal a northward shift of TC records associated with higher precipitation in the subtropical East Asia and lower precipitation east of the Philippines. These anomalies are accompanied by broad warming in the tropics and intensification of subtropical high (not shown) but limited changes over land. The tropical warming has an El Niño-like pattern though with less zonal contrast in equatorial temperature and statistically insignificant East Pacific signals. These large-scale patterns differ from those associated with the shift of DoY of TC records (Figure 4), suggesting that changes in the spatial-temporal distributions of TC activity can be associated with distinct climate processes.

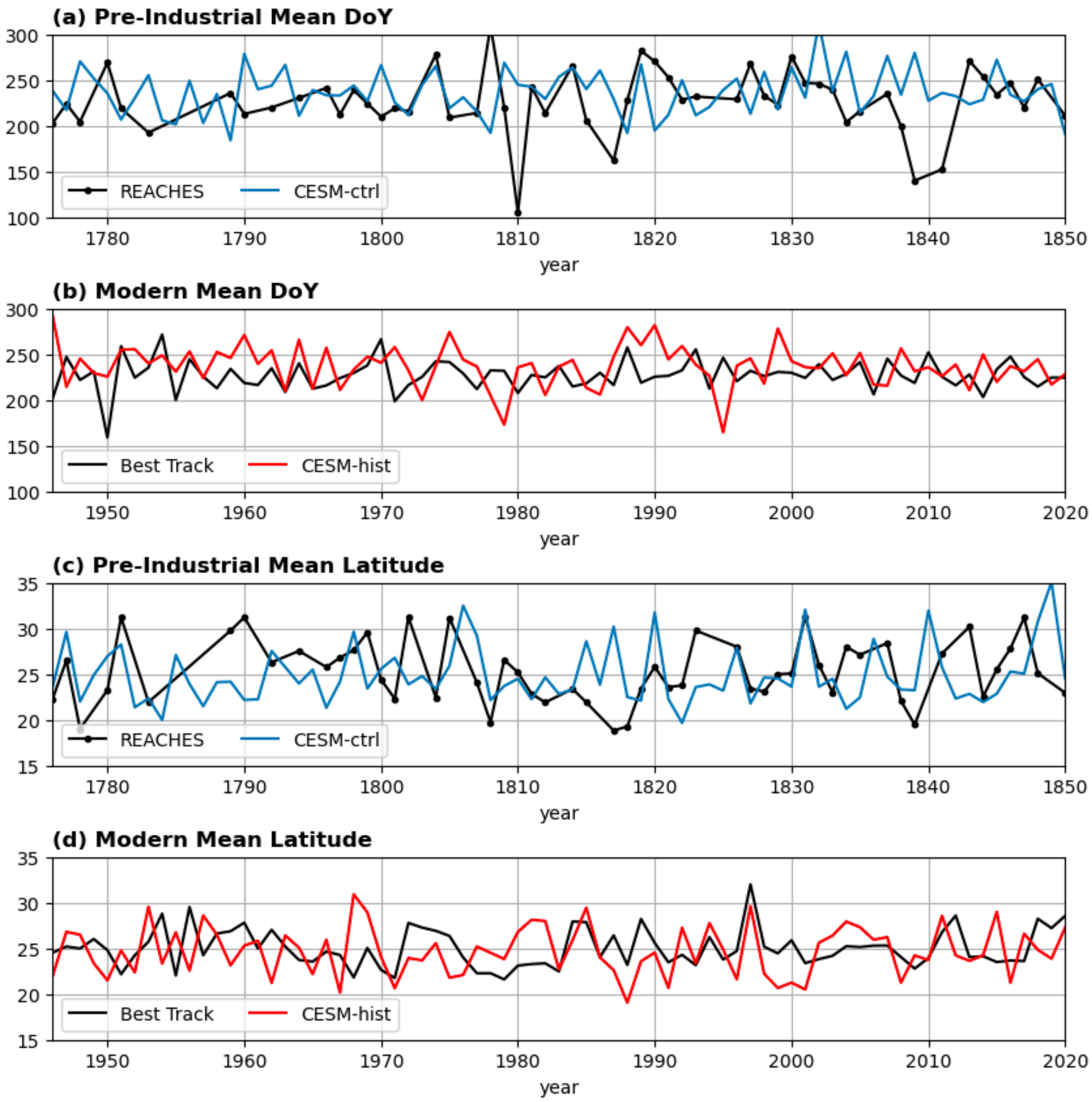

Supplementary Figure 1 *The evolution of Day of Year (DoY) and latitude metrics of TC landfall records in East Asia. The metrics consider all available samples over land and are regional, annual means. (a) Pre-industrial DoY from REACHES records and CESM control simulation. The year assignment of the CESM control simulation is arbitrary since it does not account for historical climate forcing (e.g., volcano eruptions). (b) Contemporary-era DoY from the best track observations and the CESM historical simulation. (c) Same as (a), but for the latitude metric. (d) Same as (b), but for the latitude metric. REACHES has some gap years without any TC records, and the dots in (a)(c) denote the years with TC records.*

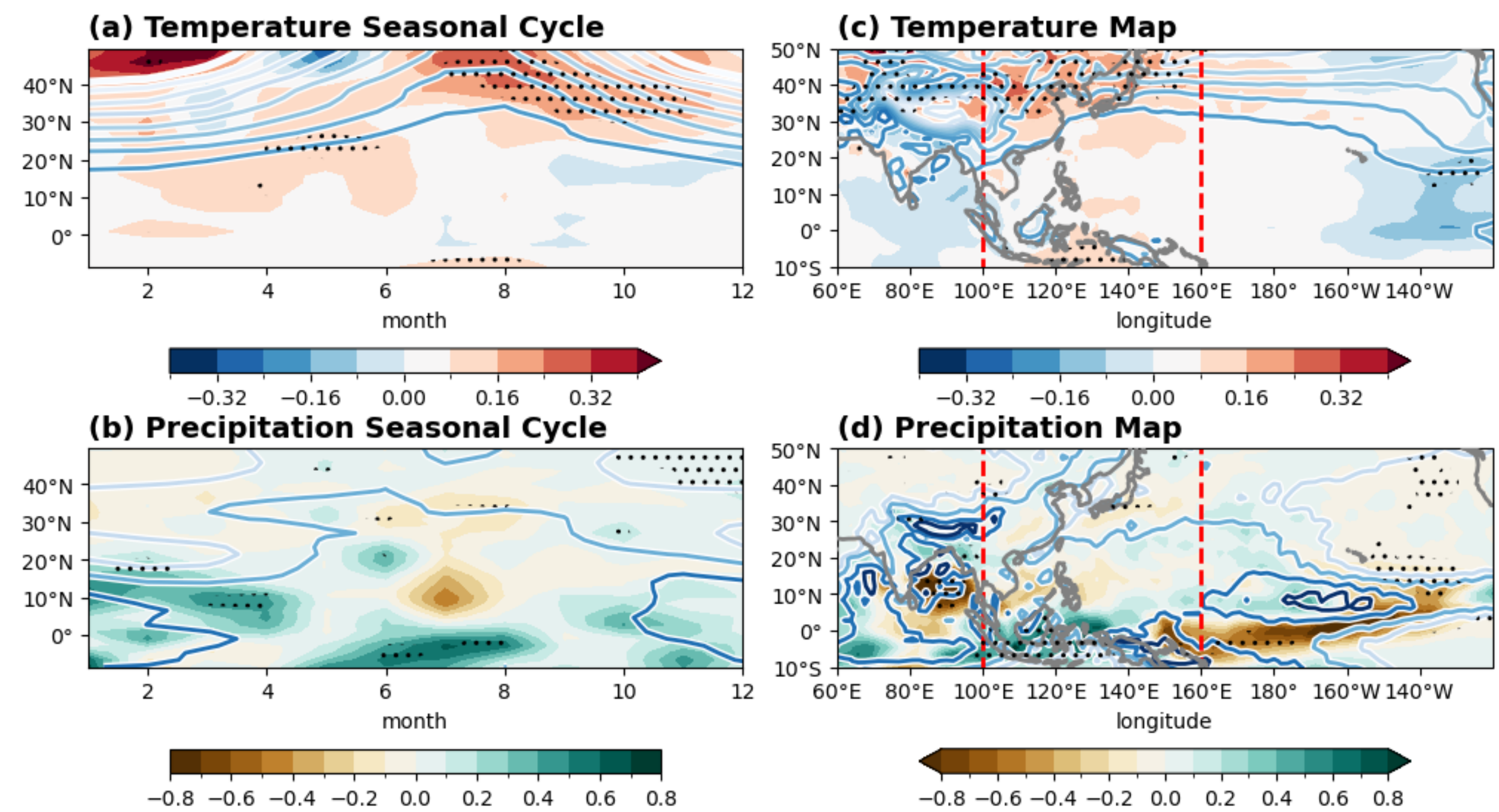

Supplementary Figure 2 *The large-scale environment differences between late-DoY and early-DoY years in 1651–1900. Based on the ranking of yearly DoY of REACHES TC records, the late-DoY and early-DoY are defined as the top and bottom 10-percentile years, respectively. The annual mean DoY is raw yearly data without any smoothing. The large-scale environment variables are from the ModE-RA. (a) Differences with the monthly mean, zonal mean surface air temperature (K) during 1651–1900. The precipitation climatology (1901–2000) is denoted by contour lines. The range of zonal averaging is 100ºE–160ºE. (b) Same as (a), but for the precipitation (shading; mm day$^{-1}$). (c) Differences between the June–October means of the surface air temperature (shading). The climatology (1901–2000) is denoted with contour lines. The red dashed lines highlight the zonal averaging range in (a). (d) Same as (c), but for the precipitation. The stippling highlights the signals above the 90% confidence level based on Student's t-test.*

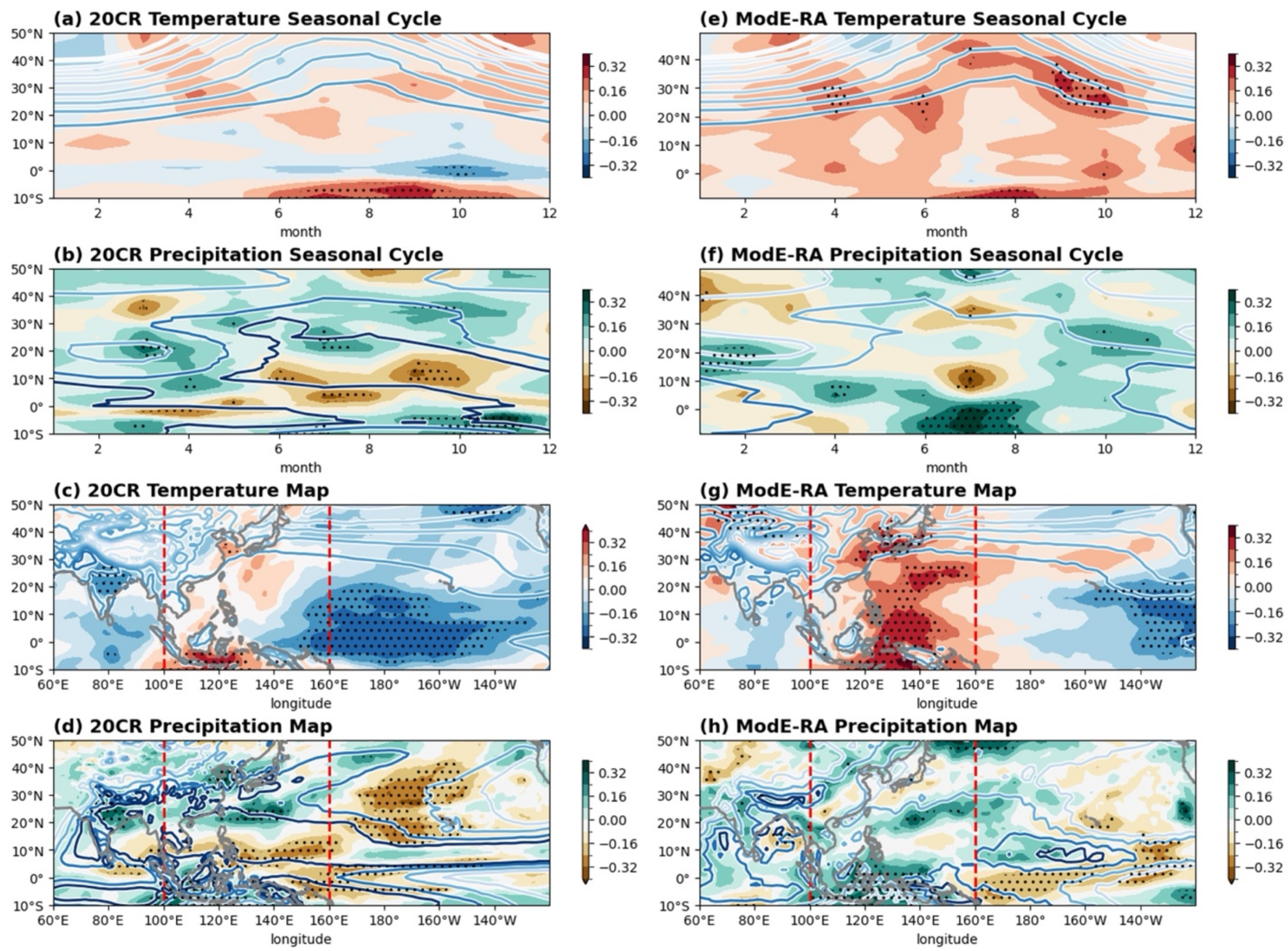

Supplementary Figure 3 *Correlations between the DoY and large-scale environment variables during 1811–1900. The yearly DoY metric is calculated with the REACHES records. In (a–d), the large-scale environment variables are from the twentieth century reanalysis. In (e–h), the large-scale environment variables are from the ModE-RA. The other figure settings are the same as in Figure 4, except that the contour levels of correlation coefficients are adjusted for clarity.*

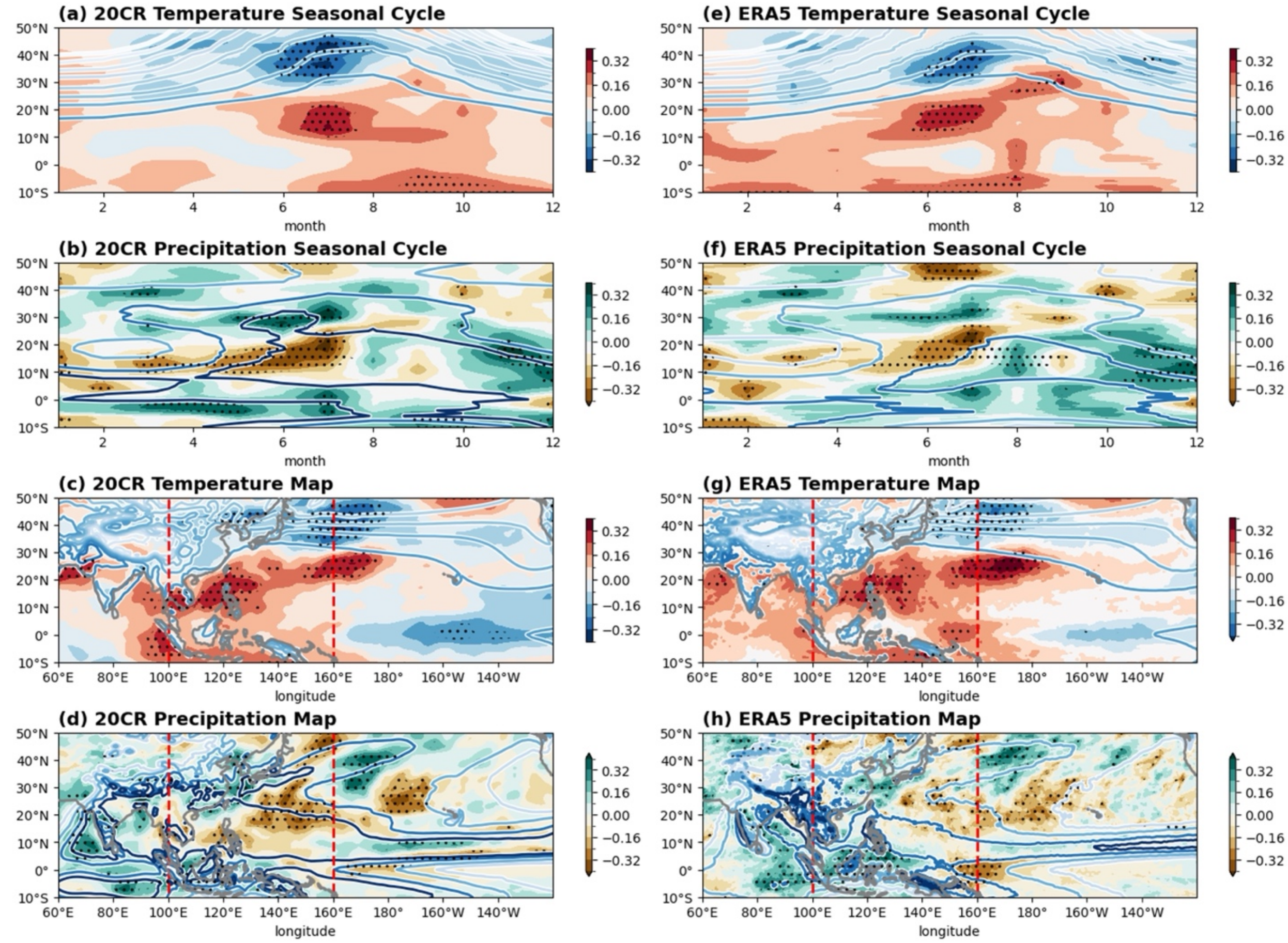

Supplementary Figure 4 *Correlations between the DoY and large-scale environment variables during 1946–2010. The yearly DoY metric is calculated with the contemporary-era best track records. In (a–d), the large-scale environment variables are from the twentieth century reanalysis. In (e–h), the large-scale environment variables are from the ERA5. The climatology of precipitation and temperature are calculated using data of 1981–2010. The other figure settings are the same as in Figure 4, except that the contour levels of correlation coefficients are adjusted for clarity.*

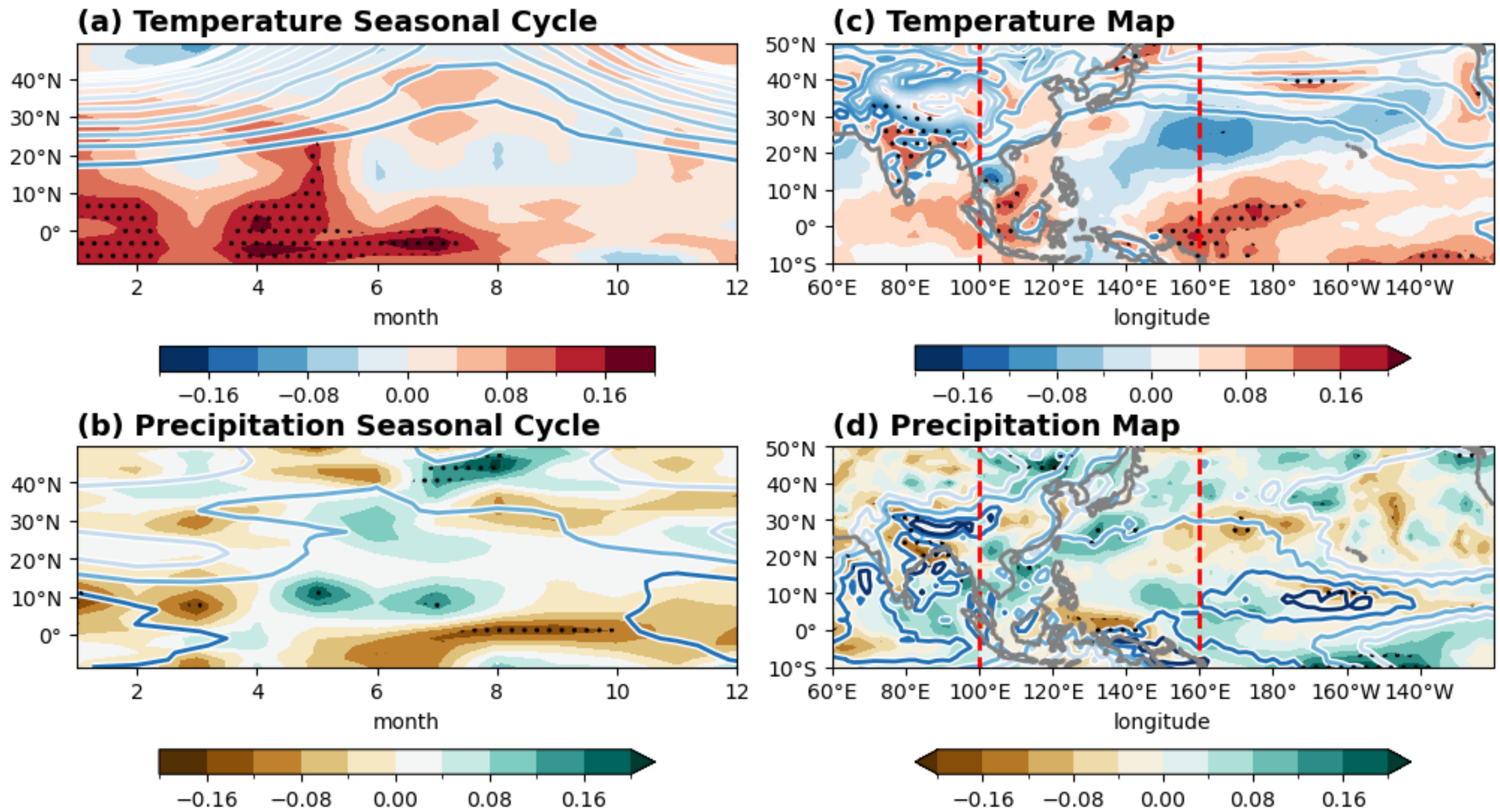

Supplementary Figure 5 *Correlations between the annual mean of latitudes of the REACHES TC records and the large-scale environment variables of the ModE-RA (1651–1900). The other figure settings are the same as in Figure 4.*

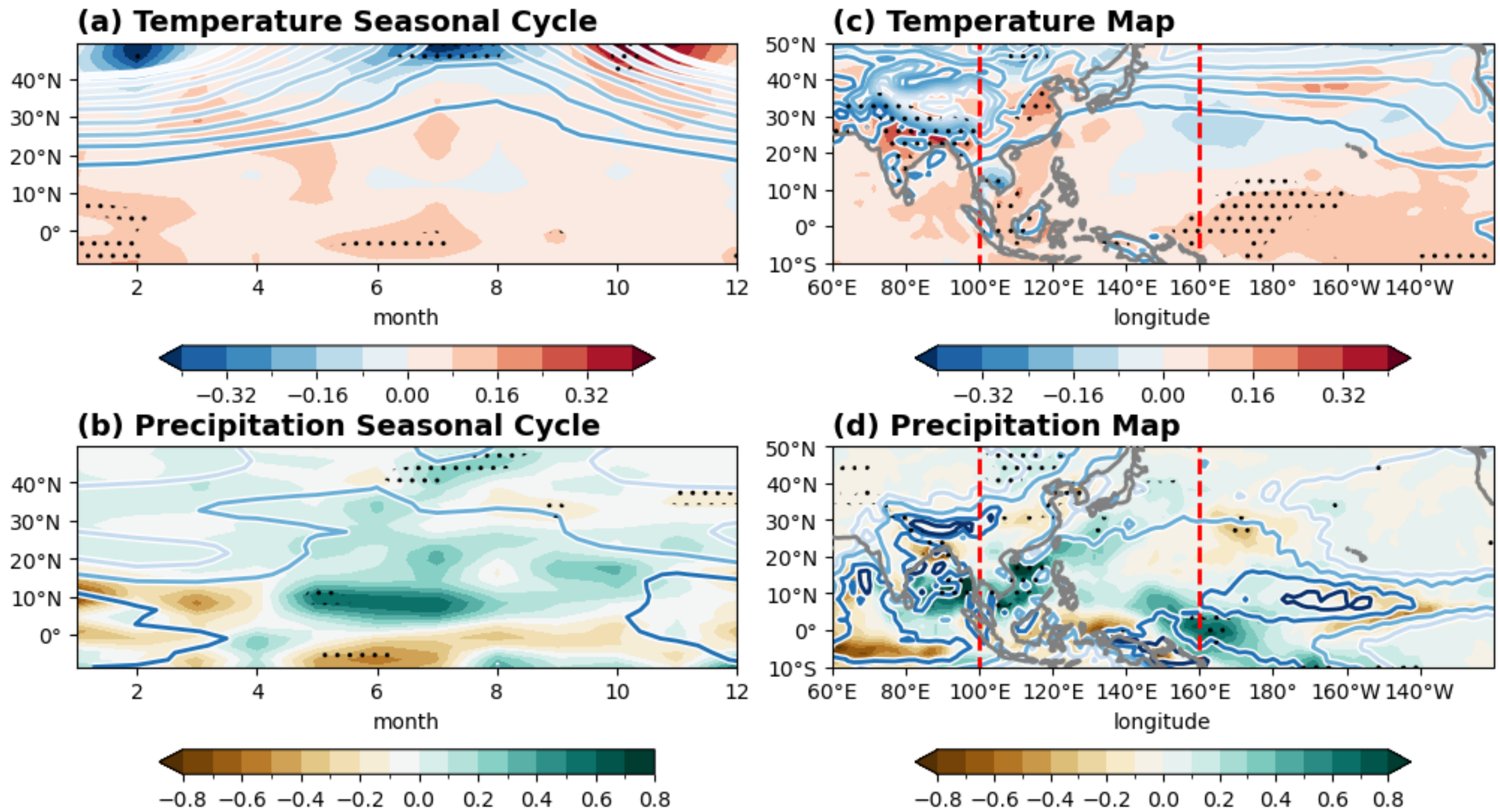

Supplementary Figure 6 *The large-scale environment differences between high-latitude and low-latitude years in 1651–1900. Based on the ranking of yearly mean latitudes of REACHES TC records, the high-latitude and low-latitude are defined as the top and bottom 10-percentile years, respectively. The other settings are similar to Supplementary Figure 2.*